# On Ordinal Covering of Proposals Using Balanced Incomplete Block Designs[1]

A. Yavuz Oruç
Department of Electrical and Computer Engineering
University of Maryland, College Park, MD 20742

Abdullah Atmaca
Department of Computer Science
Bilkent University, Ankara, Turkey

**Abstract**
A frequently encountered problem in peer review systems is to facilitate pairwise comparisons of a given set of proposals by as few as referees as possible. In [8], it was shown that, if each referee is assigned to review $k$ proposals then $\lceil n(n-1)/k(k-1) \rceil$ referees are necessary and $\lceil n(2n-k)/k^2 \rceil$ referees are sufficient to cover all $n(n-1)/2$ pairs of $n$ proposals. While the upper bound remains within a factor of 2 of the lower bound, it becomes relatively large for small values of $k$ and the ratio of the upper bound to the lower bound is not less than 3/2 when $2 \leq k \leq n/2$. In this paper, we show that, if $\sqrt{n} \leq k \leq n/2$ then the upper and lower bounds can be made closer in that their ratio never exceeds 3/2. This is accomplished by a new method that assigns proposals to referees using a particular family of balanced incomplete block designs. Specifically, the new method uses $\lceil n(n+k)/k^2 \rceil$ referees when $n/k$ is a prime power, $n$ divides $k^2$, and $\sqrt{n} \leq k \leq n/2$. Comparing this new upper bound to the one given in [8] shows that the new upper bound approaches the lower bound as $k$ tends to $\sqrt{n}$ whereas the upper bound in [8] approaches the lower bound as $k$ tends to $n$. Therefore, the new method given here when combined together with the one in [8] provides an assignment whose upper bound referee complexity always remains within a factor of 3/2 of the lower bound when $\sqrt{n} \leq k \leq n$, thereby improving upon the assignment described in [8]. Furthermore, the new method provides a minimal covering, i.e., it uses the minimum number of referees possible when $k = \sqrt{n}$ and $k$ is a prime power.

## 1. Introduction

Ordinal or relative evaluations of proposals have been suggested as a more reliable alternative in identifying high quality proposals over cardinal evaluations such as using average scores assigned to the proposals by a set of referees [1-5]. Ordinal and cardinal strengths of preferences have also been advocated in [7,9] as natural extensions of ordinal comparison models. A set covering integer programming approach was introduced in [4] to obtain as many comparisons as possible between the proposals reviewed by a fixed set of referees.

This paper is concerned with the ordinal covering of a set of proposals, i.e., assignments of proposals to referees in such a way that each pair of proposals is covered by one or more referees. Let $r(n)$ denote the number of referees used in an ordinal covering of $n$ proposals. Such a covering is called *minimal ordinal covering* with a referee capacity of $k$ if $r(n)k(k-1)/n(n-1)$ tends to 1 as $n$ tends to $\infty$ and (b) each referee reviews no more than $k$ proposals. Here, the underlying assumption is that referees can compare and rank the proposals to which they are assigned. Thus, if proposals *a, b*, and *c* are assigned to a referee then it is assumed that the pairs of proposals *ab*, *ac*, and *bc* are covered by that referee. For $n$ proposals, the problem is then to cover all $n(n-1)/2$ pairs of proposals such that each referee can review no more than $k$ proposals and that the number of referees approaches $n(n-1)/k(k-1)$ for large $n$.

---

[1] This research is funded in part by the Scientific and Technological Research Council of Turkey under grant No: 109M149. This work has been submitted to the Information Processing Letters for publication. Copyright may be transferred without notice, after which this version may no longer be accessible.

It was shown in [8] the lower bound of $\lceil n(n-1)/k(k-1) \rceil$ referees is tight for $k = n/2$, and almost tight for $k = n/3$ (11 versus 12 referees), and $k = n/4$ (18 versus 20 referees). It was further established that $\lceil n(2n-k)/k(k-1) \rceil$ referees are sufficient to generate all pairs of $n$ proposals under the same capacity constraint of $k$ proposals per referee for any $k$ that divides $n$. The main contribution of this paper is a new assignment that covers all pairs of $n$ proposals using $\lceil n(n+k)/k^2 \rceil$ referees, each with a capacity of $k$, whenever $n/k$ is a prime power, $n$ divides $k^2$, and $\sqrt{n} \leq k \leq n/2$. The new assignment relies on a balanced incomplete block design and mutually orthogonal set of Latin squares and results in a referee complexity that improves the referee complexity of the assignment described in [8] when $\sqrt{n} \leq k \leq n/2$. Furthermore, it provides a minimal ordinal covering when $k = \sqrt{n}$.

The rest of the paper is organized as follows. In Section 2, we review the block design and Latin square concepts needed in the rest of the paper. In Section 3, we present the new assignment and prove that it covers all pairs of $n$ proposals with $\lceil n(n+k)/k^2 \rceil$ referees, each with a capacity of $k$. The paper is concluded in Section 4 with the comparison of the new assignment with that presented in [8] and discussion of remaining problems.

**2. Preliminaries**

A block design is a pair $(G,B)$ where $G = \{G_1, G_2, ..., G_v\}$ is a set of $v$ elements, and $R = \{R_1, R_2, ..., R_b\}$ is a collection of $b$ subsets of $G$, called blocks[2]. For example, if $G = \{G_1, G_2, G_3, G_4, G_5, G_6\}$, then $\{(G_1, G_3, G_5), (G_2, G_4, G_5, G_6)\}$ is a block design with the two blocks, $(G_1, G_3, G_5)$ and $(G_2, G_4, G_5, G_6)$. A block design is called balanced if all blocks are of equal size and all the pairs of elements of $G$ occur in all of the blocks an equal number of times. It is called incomplete if the number of elements in every block is less than $v$. Let $\lambda$, $t$, and $r$ be positive integers, where $2 \leq t < v$. A block design $(G,B)$ is called a $(v,b,r,t,\lambda)$-balanced and incomplete block design (BIBD) if (1) $|G| = v$, $|B| = b$, (2) each element in $G$ appears in exactly $r$ blocks, (3) all blocks in $B$ have $t$ elements, and (4) each pair of elements in $G$ appears in exactly $\lambda$ blocks. We have $vr = bt$ since each of the $v$ elements in $G$ appears $r$ times in all the blocks and the union of the blocks as a multiset contains exactly $bt$ elements. It can further be shown that $\lambda(v-1) = r(t-1)$. Solving for $b$ and $r$ in terms of $\lambda$ and $t$, we have

$$b = \frac{v(v-1)\lambda}{t(t-1)}, \qquad r = \frac{(v-1)\lambda}{t-1}$$

and thus a $(v,b,r,t,\lambda)$-BIBD design is often referred to as a $(v,t,\lambda)$-BIBD as will be done here as well. In this paper, we will be concerned with BIBDs with $\lambda = 1$.

---

[2] In this paper, $G_1, G_2, ..., G_v$ will implicitly represent subsets of proposals and $R_1, R_2, ..., R_b$ will implicitly represent referees.

**Proposition 1:** Let $G = \{G_i: 1 \leq i \leq 9\}$ and

$B = \{(G_1, G_2, G_3), (G_4, G_5, G_6), (G_7, G_8, G_9), (G_1, G_4, G_7), (G_1, G_5, G_8), (G_1, G_6, G_9),$
$(G_2, G_4, G_9), (G_2, G_5, G_7), (G_2, G_6, G_8), (G_3, G_4, G_8), (G_3, G_5, G_9), (G_3, G_6, G_7)\}.$

$(G,B)$ is a $(9,3,1)$-BIBD design with each $G_i$ appearing in exactly 4 of 12 blocks.

**Proof:** It follows from a direct inspection of the blocks.

**Proposition 2:** Let $G = \{G_i: 1 \leq i \leq 16\}$ and

$B = \{(G_1,G_2,G_3,G_4),(G_5,G_6,G_7,G_8),(G_9,G_{10},G_{11},G_{12}),(G_{13},G_{14},G_{15},G_{16}),$
$(G_1,G_5,G_9,G_{13}),(G_1,G_6,G_{11},G_{16}),(G_1,G_7,G_{12},G_{14}),(G_1,G_8,G_{10},G_{15}),$
$(G_2,G_6,G_{10},G_{14}),(G_2,G_5,G_{12},G_{15}),(G_2,G_8,G_{11},G_{13}),(G_2,G_7,G_9,G_{16}),$
$(G_3,G_7,G_{11},G_{15}),(G_3,G_5,G_{10},G_{16}),(G_1,G_6,G_{12},G_{13}),(G_3,G_8,G_9,G_{14}),$
$(G_4,G_8,G_{12},G_{16}),(G_4,G_5,G_{11},G_{14}),(G_4,G_7,G_{10},G_{13}),(G_4,G_6,G_9,G_{15})\}.$

$(G,B)$ is a $(16,4,1)$-BIBD with each $G_i$ appearing in exactly 5 of 20 blocks:

**Proof:** It follows from a direct inspection of the blocks.

Certain BIBDs can be constructed using Latin squares and this fact will play a critical role in the sequel. A Latin square of order $q$ is an $q \times q$ matrix in which each row and each column is a permutation of a set of $q$ symbols [6]. Two Latin squares $L_1$ and $L_2$ are said to be orthogonal if the matrix obtained by juxtaposing $L_1$ and $L_2$ entry by entry contains each of the possible $q^2$ ordered pairs exactly once. For example, the following Latin squares are orthogonal.

Orthogonal Latin squares: $\begin{bmatrix} 1 & 2 & 3 \\ 2 & 3 & 1 \\ 3 & 1 & 2 \end{bmatrix} \times \begin{bmatrix} 4 & 5 & 6 \\ 6 & 4 & 5 \\ 5 & 6 & 4 \end{bmatrix} = \begin{bmatrix} (1,4) & (2,5) & (3,6) \\ (2,6) & (3,4) & (1,5) \\ (3,5) & (1,6) & (2,4) \end{bmatrix}$

It should be noted that not all Latin squares are orthogonal as illustrated by the following example.

Non-orthogonal Latin squares: $\begin{bmatrix} 1 & 2 & 3 \\ 2 & 3 & 1 \\ 3 & 1 & 2 \end{bmatrix} \times \begin{bmatrix} 6 & 5 & 4 \\ 5 & 4 & 6 \\ 4 & 6 & 5 \end{bmatrix} = \begin{bmatrix} (1,6) & (2,5) & (3,4) \\ (2,5) & (3,4) & (1,6) \\ (3,4) & (1,6) & (2,5) \end{bmatrix}$

A set of Latin squares, $L_1, L_2, ..., L_m$, is said to be mutually orthogonal if each pair of Latin squares is orthogonal, i.e., $L_i$ and $L_j$ are orthogonal, $1 \leq i \neq j \leq m$. We call such a set of Latin squares mutually orthogonal Latin squares and denote it by MOLS. It is known that the maximum number of Latin squares of order $q$ that can be mutually orthogonal to one another cannot exceed $q$-1. Accordingly, a set of mutually orthogonal Latin squares, $L_1, L_2, ..., L_{q-1}$ is said to be a complete MOLS, and the $q \times q$ matrix obtained by juxtaposing them entry by entry will be denoted by $L_1 \times L_2 \times ... \times L_{q-1}$.

A number of methods are known for constructing a complete set of $q$-1 MOLS of order $q$. We state the following theorem [6] without a proof:

**Theorem 1:** If $q$ is a prime power, a complete set of $q$-1 MOLS can be found by taking the nonzero elements of a finite field of order $q$, and setting the entry in the $x^{th}$ row and $y^{th}$ column in the $a^{th}$ Latin square to $f_a(x, y) = ax + y \pmod{q}$, $1 \leq a \leq q$.

**Example 1:** For $n = 3$, we compute the Latin squares as follows[3]:

$f_1(1, 1) = 2, f_1(1, 2) = 3, f_1(1, 3) = 1$
$f_1(2, 1) = 3, f_1(2, 2) = 1, f_1(2, 3) = 2$     $L_1 = \begin{bmatrix} 2 & 3 & 1 \\ 3 & 1 & 2 \\ 1 & 2 & 3 \end{bmatrix}$
$f_1(3, 1) = 1, f_1(3, 2) = 2, f_1(3, 3) = 3$

$f_2(1, 1) = 3, f_2(1, 2) = 1, f_2(1, 3) = 2$
$f_2(2, 1) = 2, f_2(2, 2) = 3, f_2(2, 3) = 1$     $L_2 = \begin{bmatrix} 3 & 1 & 2 \\ 2 & 3 & 1 \\ 1 & 2 & 3 \end{bmatrix}$
$f_2(3, 1) = 1, f_2(3, 2) = 2, f_2(3, 3) = 3$

## 3. Covering Pairs of Proposals With Balanced Incomplete Block Designs

The (9,3,1) and (16,4,1) BIBDs described in Propositions 1 and 2 above can be used to obtain a covering of $n$ proposals with 12 referees, each with a of capacity of $n/3$ and with 20 referees, each with a capacity of $k = n/4$. Here, we will only describe how such a covering is obtained for $k = n/3$ since the case of $k = n/4$ is very similar.

Divide the set of $n$ proposals into 9 subsets of $n/9$ proposals[4], and call them $G_i$, $1 \leq i \leq 9$. Identify these subsets of proposals with the elements of $G$ in the (9,3,1)-BIBD as described in Proposition 1, and let $R_j$ denote the $j^{th}$ block in this (9,3,1)-BIBD, $1 \leq j \leq 12$. Assign the $n/3$ proposals in the union of the subsets of proposals in block $R_j$ to the $j^{th}$ referee, $1 \leq j \leq 12$. To prove that all $n(n-1)/2$ pairs of proposals are covered by the 12 referees, we note that the referees assigned to blocks $R_1$, $R_2$, and $R_3$ cover the pairs of proposals in $G_i$, $1 \leq i \leq 9$. The remaining pairs of proposals belong to the products of sets of proposals, $G_i \times G_j$, $1 \leq i \neq j \leq 9$. Given that each pair $(G_i, G_j)$ $1 \leq i \neq j \leq 9$ appears in one of the blocks in the (9,3,1)-BIBD design, all of remaining pairs of proposals are covered by one at least one of the remaining 9 referees.

---

[3] Note that $0 \equiv 3 \pmod 3$.

[4] Here, we assume that $k^2 = n^2/9$ is divisible by $n$, i.e., $k^2/n = n/9$ is an integer.

If $n = 9$, i.e., $k = n/k = 3$ then the (9,3,1)-design results in a minimal ordinal covering of 9 proposals as illustrated by the following example:

**Example 2:** Let

$G_1=\{p_1\}$, $G_2=\{p_2\}$, $G_3=\{p_3\}$, $G_4=\{p_4\}$, $G_5=\{p_5\}$, $G_6=\{p_6,\}$, $G_7=\{p_7\}$, $G_8=\{p_8\}$, $G_9=\{p_9\}$.

The assignment below covers all 36 pairs of 9 proposals with 12 referees with each assigned to review three proposals. The number of referees in this assignment matches the lower bound $\lceil n(n-1)/k(k-1) \rceil = 9(9-1)/3(3-1) = 12$. ∥

|          |       |       |       |       |       |       |       |       |       |
|----------|-------|-------|-------|-------|-------|-------|-------|-------|-------|
| $r_1$    | $p_1$ | $p_2$ | $p_3$ |       |       |       |       |       |       |
| $r_2$    |       |       |       | $p_4$ | $p_5$ | $p_6$ |       |       |       |
| $r_3$    |       |       |       |       |       |       | $p_7$ | $p_8$ | $p_9$ |
| $r_4$    | $p_1$ |       |       | $p_4$ |       |       | $p_7$ |       |       |
| $r_5$    | $p_1$ |       |       |       | $p_5$ |       |       | $p_8$ |       |
| $r_6$    | $p_1$ |       |       |       |       | $p_6$ |       |       | $p_9$ |
| $r_7$    |       | $p_2$ |       | $p_4$ |       |       |       |       | $p_9$ |
| $r_8$    |       | $p_2$ |       |       | $p_5$ |       | $p_7$ |       |       |
| $r_9$    |       | $p_2$ |       |       |       | $p_6$ |       | $p_8$ |       |
| $r_{10}$ |       |       | $p_3$ | $p_4$ |       |       |       | $p_8$ |       |
| $r_{11}$ |       |       | $p_3$ |       | $p_5$ |       |       |       | $p_9$ |
| $r_{12}$ |       |       | $p_3$ |       |       | $p_6$ | $p_7$ |       |       |

Table 1. Assignment of 9 proposals to 12 referees, each with a capacity of 3.

It should be emphasized that the (9,3,1)-BIBD can be used to obtain a covering of $n$ proposals by 12 referees, each with a capacity of $k$, for any $n$ that is divisible by 9 and where $k = n/3$. The assignment in Table 2 uses 12 referees, each with a capacity of 9 to cover all 351 pairs of 27 proposals. Likewise, the same (9,3,1)-BIBD can be used to cover all 1431 pairs of 54 proposals using 12 referees, each with a capacity of 18. However, for $n > 9$, the lower bound on the number of referees becomes[8]:

$$\lceil n(n-1)/k(k-1) \rceil = \lceil n(n-1)/n/3(n/3-1) \rceil = \lceil 9(n-1)/(n-3) \rceil \geq 10$$

and hence it cannot be said that either of these assignments generated by the (9,3,1)-BIBD is a minimal ordinal covering.

| | | | | | | | | | |
|---|---|---|---|---|---|---|---|---|---|
| $r_1$ | $p_1p_2p_3$ | $p_4p_5p_6$ | $p_7p_8p_9$ | | | | | | |
| $r_2$ | | | | $p_{10}p_{11}p_{12}$ | $p_{13}p_{14}p_{15}$ | $p_{16}p_{17}p_{18}$ | | | |
| $r_3$ | | | | | | | $p_{19}p_{20}p_{21}$ | $p_{22}p_{23}p_{24}$ | $p_{25}p_{26}p_{27}$ |
| $r_4$ | $p_1p_2p_3$ | | | $p_{10}p_{11}p_{12}$ | | | $p_{19}p_{20}p_{21}$ | | |
| $r_5$ | $p_1p_2p_3$ | | | | $p_{13}p_{14}p_{15}$ | | | $p_{22}p_{23}p_{24}$ | |
| $r_6$ | $p_1p_2p_3$ | | | | | $p_{16}p_{17}p_{18}$ | | | $p_{25}p_{26}p_{27}$ |
| $r_7$ | | $p_4p_5p_6$ | | $p_{10}p_{11}p_{12}$ | | | | | $p_{25}p_{26}p_{27}$ |
| $r_8$ | | $p_4p_5p_6$ | | | $p_{13}p_{14}p_{15}$ | | $p_{19}p_{20}p_{21}$ | | |
| $r_9$ | | $p_4p_5p_6$ | | | | $p_{16}p_{17}p_{18}$ | | $p_{22}p_{23}p_{24}$ | |
| $r_{10}$ | | | $p_7p_8p_9$ | $p_{10}p_{11}p_{12}$ | | | | $p_{22}p_{23}p_{24}$ | |
| $r_{11}$ | | | $p_7p_8p_9$ | | $p_{13}p_{14}p_{15}$ | | | | $p_{25}p_{26}p_{27}$ |
| $r_{12}$ | | | $p_7p_8p_9$ | | | $p_{16}p_{17}p_{18}$ | $p_{19}p_{20}p_{21}$ | | |

Table 2. Assignment of 27 proposals to 12 referees, each with a capacity of 9.

Similarly, the (16,4,1)-BIBD design results in a minimal ordinal covering of 16 proposals by 20 referees each with a capacity of 4 since $\lceil 16(16-1)/4(4-1) \rceil = 20$ but again, for $n > 16$, the assignment is no longer minimal given that

$$\lceil n(n-1)/k(k-1) \rceil = \lceil n(n-1)/n/4(n/4-1) \rceil = \lceil 16(n-1)/(n-4) \rceil \geq 17 \text{ for } n > 16.$$

Our main result is a generalization of the (9,3,1) and (16,4,1)-BIBD designs to a $(q^2,q,1)$-BIBD design for any prime power $q \geq 2$, and efficient assignments that result from this BIBD design. We note that the assignments with 12 and 20 referees were presented in [8] without using any connection to the (9,3,1) and (16,4,1)-BIBD designs described here.

**Proposition 3**: The intersection of any two columns in any given row or the intersection of any two rows in any given column in a $q \times q$ matrix obtained by juxtaposing the entries in a complete set of $q$-1 MOLS is empty.

**Proof**: The entries in any column or any row of a $q \times q$ Latin square form a permutation of the $q$ elements used to construct the Latin square. Therefore the intersections of the juxtapositions of the entries across the rows or columns of the Latin squares in the MOLS must be empty.  ∥

**Example 3:** In the MOLS below, {1,4}∩{2,5}=∅, {1,4}∩{3,6}=∅, {2,5}∩{3,6}=∅ in the first row, and it can be verified the same holds for any other row or column. ‖

$$\begin{bmatrix} 1 & 2 & 3 \\ 2 & 3 & 1 \\ 3 & 1 & 2 \end{bmatrix} \times \begin{bmatrix} 4 & 5 & 6 \\ 6 & 4 & 5 \\ 5 & 6 & 4 \end{bmatrix} = \begin{bmatrix} (1,4) & (2,5) & (3,6) \\ (2,6) & (3,4) & (1,5) \\ (3,5) & (1,6) & (2,4) \end{bmatrix}$$

**Theorem 2:** For any prime power $q$, there exists a $(q^2,q,1)$-BIBD.

**Proof:** By Theorem 1, we can construct a complete set of MOLS using $q-1$ Latin squares of order $q$, $L_i$, $1 \leq i \leq q-1$. Let $U_i=\{(i-1)q+1,(i-1)q+2,...,(i-1)q+q\}$ denote the set of elements used in $L_i$, $1 \leq i \leq q-1$, and let $U_q = \{(q-1)q+1,(q-1)q+2,...,(q-1)q+q\}$. Let $M = L_1 \times L_2 \times ... \times L_{q-1}$ denote the $q \times q$ matrix obtained by juxtaposing $L_1, L_2, ..., L_{q-1}$ as defined in Section 2. Suppose that $M$ is modified by concatenating $(q-1)q + i$ to all the columns in the $i^{th}$ row of $M$, $1 \leq i \leq q$. Denote this new matrix by $M_a$. Let the entries of matrix $M_a$ represent the blocks of a block design. Each element in $U_1 \cup U_2 \cup ... \cup U_{q-1}$ clearly appears exactly $q$ times among these blocks. Since each element in $U_q$ is inserted into the columns of a distinct row, each element in $U_q$ must also appear exactly $q$ times among the blocks. Moreover, $U_q \cap U_i = \emptyset$ for $i = 1,2,...,q-1$. Therefore, by Proposition 3, the intersection of any two rows in any given column must be empty. Furthermore, the intersection of any two columns in any given row cannot have more than one element in common. It follows that the pairs of elements that are formed by juxtaposing the elements in the blocks of matrix $M_a$ must all be distinct.

Now to complete this block design to a $(q^2,q,1)$-BIBD, it suffices to add $U_i$, $1 \leq i \leq q$ as blocks to it and note that (a) that the resulting block design consists of $q^2+q$ blocks, each comprising $q$ elements, (b) each element in $U_1 \cup U_2 \cup ... \cup U_q$ appears exactly in $q+1$ blocks, and (c) each pair of elements appears in exactly one block. ‖

**Remark 2:** The equation

$$q^2 \binom{q}{2} + q \binom{q}{2} = \binom{q^2}{2}$$

captures the fact that the set of all pairs of $q^2$ elements in $U_1 \cup U_2 \cup ... \cup U_q$ is obtained by the union of the set of all pairs of all elements in the $q^2$ blocks identified with the entries of $M_a$ plus all pairs of elements generated by the elements in $U_i$, $1 \leq i \leq q$. We also note that the (9,3,1) and (16,4,1)-BIBD designs are special cases of this $(q^2,q,1)$-BIBD construction with $q = 3$ and $q = 4$, respectively. ‖

**Example 4:** Let $q = 5$. The following four Latin squares form a complete set of MOLS of order 5.

$$L_1 = \begin{bmatrix} 1 & 2 & 3 & 4 & 5 \\ 2 & 3 & 5 & 1 & 4 \\ 3 & 5 & 4 & 2 & 1 \\ 4 & 1 & 2 & 5 & 3 \\ 5 & 4 & 1 & 3 & 2 \end{bmatrix}, L_2 = \begin{bmatrix} 6 & 7 & 8 & 9 & 10 \\ 8 & 10 & 9 & 7 & 6 \\ 9 & 6 & 7 & 10 & 8 \\ 10 & 9 & 6 & 8 & 7 \\ 7 & 8 & 10 & 6 & 9 \end{bmatrix}, L_3 = \begin{bmatrix} 11 & 12 & 13 & 14 & 15 \\ 14 & 11 & 12 & 15 & 13 \\ 15 & 14 & 11 & 13 & 12 \\ 12 & 13 & 15 & 11 & 14 \\ 13 & 15 & 14 & 12 & 11 \end{bmatrix}, L_4 = \begin{bmatrix} 16 & 17 & 18 & 19 & 20 \\ 20 & 19 & 16 & 18 & 17 \\ 17 & 18 & 20 & 16 & 19 \\ 18 & 20 & 19 & 17 & 16 \\ 19 & 16 & 17 & 20 & 18 \end{bmatrix},$$

Matrix $M$ is constructed by juxtaposing these Latin squares as follows:

$$M = \begin{bmatrix} (1,6,11,16) & (2,7,12,17) & (3,8,13,18) & (4,9,14,19) & (5,10,15,20) \\ (2,8,14,20) & (3,10,11,19) & (5,9,12,16) & (1,7,15,18) & (4,6,13,17) \\ (3,9,15,17) & (5,6,14,18) & (4,7,11,20) & (2,10,13,16) & (1,8,12,19) \\ (4,10,12,18) & (1,9,13,20) & (2,6,15,19) & (5,8,11,17) & (3,7,14,16) \\ (5,7,13,19) & (4,8,15,16) & (1,10,14,17) & (3,6,12,20) & (2,9,11,18) \end{bmatrix}$$

Matrix $M_a$ is formed by inserting 21 to the 1$^{st}$ row of $M$, 22 to its 2$^{nd}$ row, 23 to its 3$^{rd}$ row, 24 to its 4$^{th}$ row and 25 to its 5$^{th}$ row as shown below:

$$M_a = \begin{bmatrix} (1,6,11,16,21) & (2,7,12,17,21) & (3,8,13,18,21) & (4,9,14,19,21) & (5,10,15,20,21) \\ (2,8,14,20,22) & (3,10,11,19,22) & (5,9,12,16,22) & (1,7,15,18,22) & (4,6,13,17,22) \\ (3,9,15,17,23) & (5,6,14,18,23) & (4,7,11,20,23) & (2,10,13,16,24) & (1,8,12,19,24) \\ (4,10,12,18,24) & (1,9,13,20,24) & (2,6,15,19,24) & (5,8,11,17,24) & (3,7,14,16,24) \\ (5,7,13,19,25) & (4,8,15,16,25) & (1,10,14,17,25) & (3,6,12,20,25) & (2,9,11,18,25) \end{bmatrix}$$

We obtain the blocks of the (25,5,1)-BIBD by combining the entries of $M_a$ with the sets

$U_1 = \{1, 2, 3, 4, 5\}$, $U_2 = \{6, 7, 8, 9, 10\}$, $U_3 = \{11,12,13,14,15\}$,

$U_4 = \{16,17,18,19,20\}$, $U_5 = \{21, 22, 23, 24, 25\}$ as

{(1, 2, 3, 4, 5) (6, 7, 8, 9, 10) (11, 12, 13, 14, 15) (16, 17, 18, 19, 20) (21, 22, 23, 24, 25)
(1, 6, 11, 16, 21) (2, 7, 12, 17, 21) (3, 8, 13, 18, 21) (4, 9, 14, 19, 21) (5, 10, 15, 20, 21)
(2, 8, 14, 20, 22) (3, 10, 11, 19, 22) (5, 9, 12, 16, 22) (1, 7, 15, 18, 22) (4, 6, 13, 17, 22)
(3, 9, 15, 17, 23) (5, 6, 14, 18, 23) (4, 7, 11, 20, 23) (2, 10, 13, 16, 23) (1, 8, 12, 19, 23)
(4, 10, 12, 18, 24) (1, 9, 13, 20, 24) (2, 6, 15, 19, 24) (5, 8, 11, 17, 24) (3, 7, 14, 16, 24)
(5, 7, 13, 19, 25) (4, 8, 15, 16, 25) (1, 10, 14, 17, 25) (3, 6, 12, 20, 25) (2, 9, 11, 18, 25)}

As mentioned in Remark 2, this BIBD construction satisfies the equation:

$$25\binom{5}{2} + 5\binom{5}{2} = 300 = \binom{25}{2}$$

and thus generates all pairs of elements in the set $\{1,2,\ldots,25\}$.

Now to obtain a minimal ordinal covering of a set of $n$ proposals from this $(q^2,q,1)$-BIBD, we set $n = q^2$, and set the capacity of each referee, $k$ to $q$. Then the lower bound on the number of referees becomes

$$\lceil n(n-1)/k(k-1) \rceil = \lceil q^2(q^2-1)/q(q-1) \rceil = q(q+1) = q^2 + q$$

matching the upper bound on the number of referees obtained by the $(q^2,q,1)$-BIBD described in Theorem 2. In the above example, setting $n = 25$ and $k = 5$ gives 30 referees and this matches the lower bound $25(25-1)/5(5-1) = 30$.

For other values of $q$, the $(q^2,q,1)$-BIBD still produces an ordinal covering but such an assignment will no longer be minimal. In particular, for any prime power, $q = n/k$, where $\sqrt{n} \leq k \leq n/2$ and $n$ divides $k^2$, the $(q^2,q,1)$-BIBD gives an ordinal covering with $q(q+1) = n/k \, (n/k +1) = n(n+k)/k^2$ referees, each with a capacity of $k$.

## 4. Concluding Remarks

We presented a method to obtain a minimal ordinal covering of a set of $n$ proposals using an $(n,\sqrt{n},1)$-BIBD. The method works for any $n$ that is a square of a prime power and requires $n+\sqrt{n}$ referees, each with a capacity of $\sqrt{n}$. By adjusting the parameters in the BIBD design, i.e., using a $(n^2/k^2,n/k,1)$-BIBD, and requiring that $n/k$ be a prime power, the number of referees, each with a capacity of $k$, can be reduced to $\lceil (n/k)(n/k+1) \rceil$, where $n$ divides $k^2$ and $\sqrt{n} \leq k \leq n/2$ or equivalently, $2 \leq n/k \leq \sqrt{n}$. The condition that $n$ should divide $k^2$ ensures that the set of $k$ proposals assigned to each referee can be partitioned into $n/k$ sets of proposals to form the entries of the blocks in the $(n^2/k^2,n/k,1)$-BIBD. For example, if $n = 32$, $k = 8$, each of the $(n/k)(n/k+1) = 20$ referees are assigned 8 proposals that can be partitioned into $n/k = 4$ groups of 2 proposals to form the entries of the 20 blocks in a (16,4,1)-BIBD. The number of referees used is quite close to the lower bound of $\lceil n(n-1)/k(k-1) \rceil = \lceil 32 \times 31/8 \times 7 \rceil = 18$ referees but clearly not minimal. It is possible to remove the restriction that $n$ should divide $k^2$ for the assignment to work but doing so requires more than $\lceil (n/k)(n/k+1) \rceil$ referees. Further discussion of how this can be accomplished will be deferred to another place.

When $\sqrt{n} \leq k \leq n/2$ and $n$ divides $k^2$, the ordinal covering given in this paper results in a smaller number of referees than the ordinal covering described in [8] as can be seen from the inequality:

$$\underbrace{\frac{n(n+k)}{k^2}}_{\text{This paper's upper bound}} \leq \underbrace{\frac{n(2n-k)}{k^2}}_{\text{upper bound in [8]}} \quad \text{when } \sqrt{n} \leq k \leq n/2$$

On the other hand, when $n/2 \leq k \leq n$, the number of referees used in [8] becomes smaller that the upper bound obtained here. Furthermore, we have

$$\overbrace{\frac{n(n+k)}{k^2}}^{\text{This paper's upper bound}} \Big/ \overbrace{\frac{n(n-1)}{k(k-1)}}^{\text{lower bound in [8]}} = \frac{(n+k)(k-1)}{(n-1)k} \le 3/2 \text{ when } k \le n/2$$

$$\overbrace{\frac{n(2n-k)}{k^2}}^{\text{upper bound in [8]}} \Big/ \overbrace{\frac{n(n-1)}{k(k-1)}}^{\text{lower bound in [8]}} = \frac{(2n-k)(k-1)}{(n-1)k} \le 3/2 \text{ when } k \ge n/2$$

Therefore, the two methods can be combined together to obtain an assignment that yields a covering with a referee complexity that remains within a factor of 3/2 of the lower bound in [8].

It remains open if the assignments described in this paper and in [8] can be improved further. In particular, it is not known if there exists a minimal ordinal covering of a set of $n$ proposals with a referee capacity of $k$, for all $k \le n$. Another direction that remains unexplored is to investigate ordinal covering of proposals by referees with different capacities.